# Size dependent photoemission study by electrochemical coarsening of nanoporous gold

*Fatemeh Ebrahimi [a,b]\*, Xinyan Wu [a], Maurice Pfeiffer [a], Hagen Renner [a], Nadiia Mameka [c], Manfred Eich [a,b], Alexander Petrov [a,b]*

[a] Hamburg University of Technology – Institute of Optical and Electronic Materials – Hamburg, Germany

[b] Helmholtz-Zentrum Hereon – Institute of Functional Materials for Sustainability – Geesthacht, Germany

[c] Helmholtz-Zentrum Hereon – Institute of Materials Mechanics – Geesthacht, Germany

\*E-mail: fatemeh.ebrahimi@tuhh.de , Phone number: (+49)4042878-3896

The generation and utilization of hot charge carriers in plasmonic materials have emerged as a topic of significant importance, with profound implications across multiple disciplines, including optoelectronics, photovoltaics, photocatalysis, and sensing. In this study, we investigate the hot electron transfer from nanoporous gold (npAu) in dependence of the structure size, utilizing both the nanoscale feature size and the interconnected nature of this material. We employ photoelectron injection from nanoporous gold into the electrolyte under UV illumination as a test electron transfer process. Nanoporous gold thin films with sub-10 nm initial ligament diameter are stepwise coarsened by potential cycles in a photoelectrochemical setup, thereby allowing us to precisely probe the influence of ligament diameter on the photocurrent response. The resulting ligament diameter variations are confirmed by scanning electron microscopy (SEM) analysis. As the ligament diameter increased from 8 to 16 nm, there was a corresponding decrease in quantum efficiency proportional to the inverse ligament diameter squared. Such dependency is expected for electrons excited by surface collisions. For the small ligament diameter of 10 nm we estimate an emission efficiency of excited 6sp electrons as 3.14%, reaching 23% for the surface excited electrons.

# INTRODUCTION

Metallic nanostructures have been extensively explored due to significant absorption by excitation of free electrons [1] and possibility to utilize the absorbed energy for different applications.[2-4] Among the various geometries investigated for this purpose, nanoporous metals have emerged as promising candidates due to their distinct advantages over particles.[5-12] Specifically, nanoporous metals offer broadband absorption, electric connectivity and a high surface-to-volume ratio.[5-10,13] Strong broadband absorption results from reduced filling fraction of metal allowing incident light to penetrate deep into the medium and to be absorbed. This is an important advantage in comparison to plasmonic particles that absorb strongly only at resonance.[10] In addition, nanoporous metals such as nanoporous gold (npAu) provide excellent electrical connectivity without the need for a support electrode. The large surface-to-volume ratio is particularly advantageous for the transfer of hot electrons to the adjacent electrolyte before the hot carrier's energy has been distributed among other electrons via collisions. Moreover, nanoporous metals are a suitable option for coating with semiconductors, thereby enabling the formation of a heterojunction.[14]

Photoemission occurs through the ejection of a hot electron originating from the absorption of a photon in the metal. When subjected to ultraviolet irradiation, excited electrons in the metal have sufficient energy to be directly injected into water.[15] In this case, the electron emits into the effective conduction band of water and is afterwards solvated in the electrolyte.[9,15] These solvated electrons can further participate in redox reactions with species present in the electrolyte solution.[9] Such photoemission process was extensively studied on flat metal electrodes.[15,16] The excitation of electrons in the metal by photon absorption requires the collision of the electron with a third partner in order to satisfy conservation of momentum. Such collision can take place in the volume, with phonons or defects, or directly at the surface of metal. Thus, volume and surface photoemission are differentiated.[16-19] Notably, the generation of hot electrons through surface collisions in metals becomes increasingly significant when the characteristic size of the structures is reduced below 30 nm.[7,20,21]

Our previous investigations of thermally coarsened npAu have shown the substantial contribution of surface photoemission from ligaments smaller than 30 nm; however, showing a saturation at very small ligament diameter.[7] This saturation was unclear in nature. Thus, we conducted further studies for small ligament diameter in the 10 nm range. Herein, we avoided inaccuracies resulting from ex-situ thermal coarsening due to variations between different samples and varying locations of illumination. Here we implemented an in-situ coarsening and characterization technique which keeps the sample and location of investigation fixed.

NpAu is typically prepared through alloy corrosion or dealloying using either free corrosion processes or electrochemical dealloying techniques.[22-24] The ligament and pore size of npAu can be adjusted within the range of a few nanometers to several microns through thermal annealing[25,26] or prolonged treatment with concentrated acid.[27] However, these conventional approaches are not suitable for in-situ studies of photoemission due to the need for sample transfer, which may result in inaccuracies as discussed above. Additionally, thermal annealing, being a rapid process, is not ideal for achieving coarsening of the nanoporous structure in small steps.[28] In order to overcome these limitations, electrochemical cycling of npAu in the presence of anions offers a promising solution to further coarse the structure. In this study, we employed electrochemical cycling of npAu with an initial ligament diameter of 8 nm to intentionally coarsen the structure. By carefully adjusting the potential window, we successfully achieved precise control over coarsening steps.

Through our study, we observed a notable trend: as the ligament diameter increased from 8 to 16 nm, the quantum efficiency exhibited a corresponding decrease, following with high precision the parabolic dependence on inverse ligament diameter. Our finding supports theoretical predictions indicating that surface and geometrical modifications significantly affect plasmon decay rates and hot carrier mobilization.[29]

We did not find any saturation effect within this range of ligament diameters. The prevalence of the quadratic term over linear term in the relationship shows the dominant role of surface photoemission in the observed photocurrent with a smaller contribution of volume photoemission.

**METHODS**

**Fabrication of Nanoporous Gold Photoelectrodes:** Nanoporous gold thin films were obtained through electrochemical dealloying of white gold leaf precursors of 6 carat (Gerstendoerfer, Germany) and confirmed by EDS that the atomic percentage of gold is about 18%. The electrochemical dealloying was done in a three-electrode cell with a gold leaf served as a working electrode, a coiled silver wire (0.1 mm in diameter) as a counter electrode, and a pseudo-Ag/AgCl electrode (0.53 V vs. reversible hydrogen electrode in 1 M $HClO_4$) as a reference electrode. All potentials in this paper are converted to reversible hydrogen electrode (RHE). The electrolyte was 1 M $HClO_4$ (70% p.a., Merk) in ultrapure water (18.2 MΩ, Sartorius). A potentiostat (Autolab, Metrohm) was used in the electrochemical dealloying experiments.

A gold leaf with 30 mm × 60 mm dimensions was placed into the electrochemical cell and connected to the potentiostat via a thin gold wire. The dealloying of the leaf used a two-step potentiostatic technique. First, we set the voltage at 1.26 V until the current fell below $2\times10^{-5}$ A. This lower starting dealloying potential mitigated excessive shrinkage and cracking of the leaf. Subsequently, the voltage was raised to 1.36 V to further remove Ag. After dealloying, the leaf was transferred by a microscope glass slide into ultra-pure water to wash off all remaining electrolyte. Then, we used a 25.4 mm × 25.4 mm × 1 mm quartz slide (Ted Pella, Inc.) as a substrate. It was cleaned in acetone, dried, and exposed to airflow to remove dust. The cleaning was followed by surface functionalization in 30 mM 3-mercaptopropyltrimethoxysilane (MPTMS) in toluene for 1 hour. MPTMS acts as an adhesion-promoting molecular layer and in this process, the surface-modified substrates possessed pendent thiol groups on which the npAu thin films could be adsorbed. The surface functionalization prevented layer delamination without covering the ligament surface of npAu.[60] Subsequently, the resulting npAu thin film was carefully transferred on the surface-modified quartz slide for further photoelectrochemical measurements.

For microstructural characterization, scanning electron microscopy was performed using the Zeiss Supra VP55 microscope with a 10 kV accelerating voltage in in-lens mode. The working distance was approximately $2 \pm 0.5$ mm, and the aperture was set to 10 µm.

**Electrochemical and Photoelectrochemical Measurements:** A npAu electrode consisting of a 140 nm-thick leaf on a quartz substrate was positioned at the backside of a photoelectrochemical cell equipped with a transparent window for illumination. The surrounding area of the npAu layer was firmly pressed against a conductive foam connected to the working electrode on the backside of the cell. A pseudo Ag/AgCl and a platinum wire were employed as reference electrode and counter electrode, respectively. Before the measurement, 0.5M $H_2SO_4$ electrolyte was purged with Ar (99.999 % purity) for 30 minutes to remove any dissolved gases. To ensure consistency in potential references, the potentials were rescaled versus the reversible hydrogen electrode (0.53 vs. RHE) in 0.5M $H_2SO_4$.

Photocurrent measurements were conducted with a commercially available setup from Zahner Elektrik, Zennium electrochemical workstation coupled with controlled intensity-modulated photospectroscopy (Zahner P211) in PECC-2 cell. The light source was tunable light (TLS03). For photoemission measurement the UV source with 291nm wavelength was selected. The on total power of the illumination was 0.52 mW within the illuminated area of 28.26 mm$^2$, so that the intensity of light was 0.184W/cm², determined within the equal geometric setup as used in the photoelectrochemical cell and placing the photodetector at the same position as the investigated sample.

**UV-VIS Measurements:** The optical absorption spectra of the samples were measured using a UV-vis spectrometer (Perkin Elmer Lambda 1050). In order to prevent the absorption of UV radiation by the glass substrate, the nanoporous gold samples were transferred onto fused silica substrates, which are commonly employed in photoelectrocatalytic investigations. The samples were immersed in a 0.5 M $H_2SO_4$ electrolyte, and any excess liquid was carefully removed before conducting spectroscopy measurements. Using integrating sphere module, the absorption spectra were determined by subtracting the contributions of the diffuse transmission and reflection from the total incident light intensity expressed as a percentage.

## RESULTS AND DISCUSSION

**Coarsening of Nanoporous Gold Upon Potential Cycling**

In this study, we focus on free-standing thin films of npAu. The samples were fabricated from a thick silver-gold leaf utilizing the electrochemical dealloying method in an acid solution.[30] This resulted an npAu film with a ligament diameter below 10 nm, as depicted in Fig. 1(a); the fabrication details can be found in the experimental section. These thin films were utilized in a photoelectrochemical cell as a working electrode for the subsequent study. The initial ligament diameter is smaller than 10 nm, resulting in a high surface-to-volume ratio, and, due to increases in small steps, allowing a more detailed analysis of the dependence of the photoemission on the ligament diameter as compared to the 40 nm initial ligament diameter used in the previous studies on electrochemical cycling.[31]

Our approach for increasing ligament diameter of the npAu structure is based on an electrochemical coarsening process, which involves a repeatable gold oxide layer formation and its reduction through the electrochemical cycling of npAu in an acid solution by a cyclic voltammetry (CV). We employed 0.5M $H_2SO_4$ electrolyte and a suitable electrochemical potential window to achieve coarsening of the npAu thin film electrodes.

Usually, the cyclic voltammetry (CV) of gold exhibits a current peak associated with a surface oxide layer formation between 1.4 and 1.6 V vs. RHE, which is followed by a minimum at approximately 1.65 V, referred to as the Burshtein minimum.[32] Exceeding the Burshtein minimum leads to the onset of the oxygen evolution reaction. At these higher potentials and at prolonged times, a place exchange of Au and O leads to the further oxidation of the surface and the formation of 3D bulk oxide in the form of $Au_2O_3$ or hydrated $Au_2O_3$.[33] After the subsequent reduction, this further roughens the surface.[34]

In order to preserve the characteristic morphology of npAu and prevent gold etching, we avoided potential windows close to the oxygen evolution reaction.[28] Consequently, the electrochemical coarsening of npAu was first achieved within the potential range of 0.83-1.63V vs. vs. reversible hydrogen electrode (RHE) (see Supplementary information Note 1). In order to better control the coarsening process and thus to obtain a smaller increase of the feature size per a CV cycle, we then limited the upper vertex potential by the first oxidation peak of 1.43 V. As can be seen in Fig. 1 (b), the npAu electrode went through 15 CV scans with a scan rate of 10 mV/s within the potential range of 0.83-1.43V vs. RHE. Scanning electron microscopy (SEM) images obtained after 15 scans of the CV verified the coarsening of the ligaments from a size average of 8.4 nm to a size average of 17.8 nm without observable changes in morphology, as can be seen in Fig.1 (a) and (c). The selected parameters resulted in a slow coarsening rate, and hence, the suitable control over the coarsening for the photoemission investigation. Coarsening the npAu in an electrochemical cell with an $H_2SO_4$ electrolyte is advantageous because the same electrolyte can also be used for in-situ photocurrent measurements or other photoelectrochemical investigations simultaneously. In our case, the $H_3O^+$ ions in the $H_2SO_4$ electrolyte play a crucial role as a charge carrier in capturing the emitted electrons upon illumination.[35] To better visualize the reaction cascade from the emitted electron to the measured current, a scheme has been included in Fig. S5.

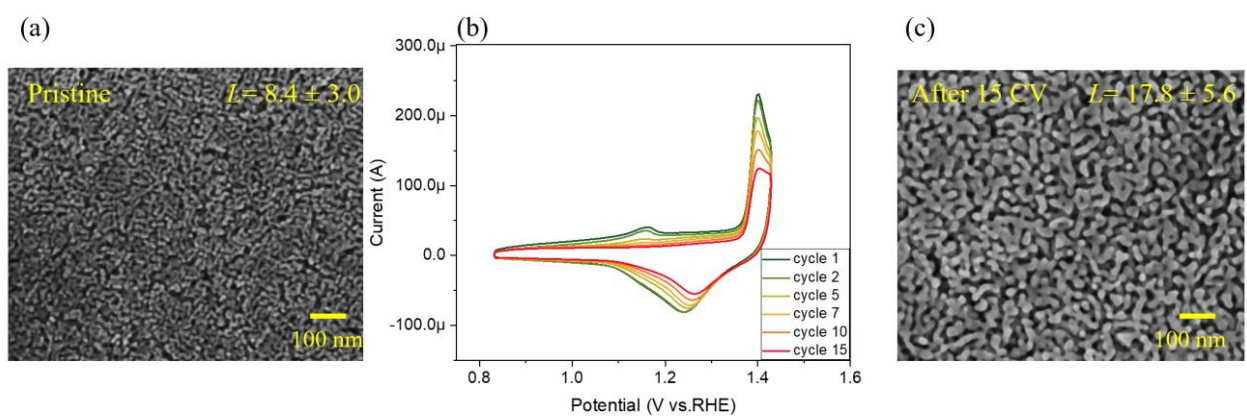

**Figure 1.** (a) Top-view SEM image of the pristine sample after dealloying. (b) Representative CV scans of npAu in the potential window of 0.53 - 1.43 V vs. RHE; the electrolyte was 0.5 M $H_2SO_4$ and the scan rate was 10 mV/s. (c) Top-view SEM image of the sample after 15 CV scans.

During CV scanning of npAu, at the potentials in the double-layer regime, the surface diffusion of gold atoms in electrolyte solutions plays a role in electrochemical coarsening. In electrolyte solutions, adsorbate interactions can affect surface diffusion parameters dramatically, leading to the enhanced surface diffusion coefficient ($D_s$) values in the order of $10^{-14}$ $cm^2$ $s^{-1}$.[36] In contrast, the $D_s$ coefficient measurements for gold in vacuum or air yield lower values in the order of $10^{-16}$ to $10^{-20}$ cm$^2$ s$^{-1}$.[37] Other factors influencing $D_s$ of gold, such as temperature, applied potential, and the presence of different adsorbed anions, are crucial for comprehending the coarsening phenomenon. Upon the electrochemical coarsening, the surface diffusion coefficient of gold atoms increases with the applied voltage, resulting in coarser ligaments. As far as there is no oxide formation on the surface, the decay process is strongly influenced by the energy barrier associated with adatom diffusion on the electrode surface. At higher electrode potentials result in greater acceleration of the island decay rate due to the increased driving force for adatom diffusion.[38]

When the cycling potential of npAu reaches the oxide-formation potential region, we observe a further coarsening of gold ligaments. The oxide is reduced in the subsequent backward cycle. Changes in the surface of flat gold during the formation of the oxide layer and a dramatic change in the surface structure after reduction were reported previously.[39] On an atomic scale, the surface was reported to be disordered or aperiodic in the oxide region.[40] We observed that by limiting the oxidation reaction, the coarsening per cycle is slowed down (see Supplementary Note 1 in the Supplementary Information for more details on coarsening in the bigger potential window).

**Optical Absorptivity of Nanoporous Gold Thin Film**

The absorptivity of 140 nm-thick npAu layer on a quartz substrate was determined using diffuse transmission and reflection measurements in a UV-Vis spectrometer equipped with an integrated sphere module. The absorptivity of the npAu film was measured before and after subjecting the sample to various number of CV scans within the potential window of 0.53 - 1.43 V vs. RHE (Fig. 2). NpAu demonstrates broadband absorbance across the entire visible wavelength range, in excellent agreement with previous reports,[41,42] This absorbance is distinct from both Au nanoparticles and bulk Au,[43] Besides the interband absorption in the wavelength range below 500 nm involving 5d electrons, npAu exhibits significant absorption of photons by free 6sp electrons at longer wavelengths. The explanation for this effect can be partly given by the effective medium model of npAu,[8] where small Au filling fraction shifts the effective plasma frequency to longer wavelengths turning the reflective metal into a lossy dielectric for a wide wavelength range. In addition, the increased surface enhances the frequency of electron collisions with the surface, facilitating photon absorption. Thus, absorptivity of npAu is the highest for the smallest ligament size. Both circumstances result in a large field penetration into the film and a strong absorbance.

Consequently, we see that absorptivity is dependent on the number of CV scans (i.e., ligament diameter), with coarser structures exhibiting the slightly reduced optical absorptivity due to decreased collision frequency of electrons in larger ligaments.

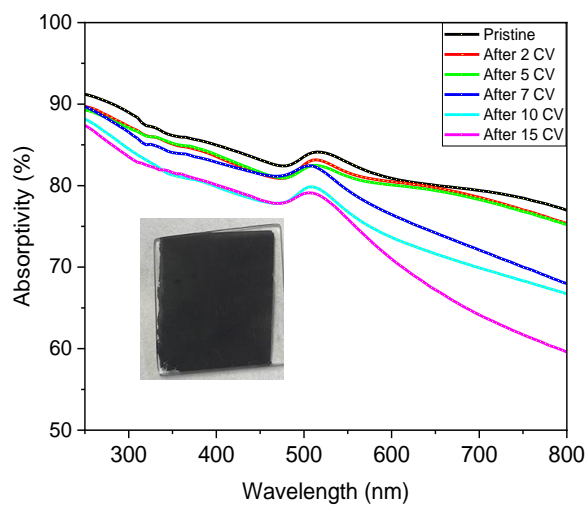

**Figure 2.** Absorptivity spectra of npAu films before and after potential cycling with different numbers of CV scans. During the measurements, the npAu films were soaked with 0.5 M $H_2SO_4$ to provide similar reflection and transmission characteristics as the npAu film used in the photoelectrochemical measurement. The insert photograph shows a typical pristine npAu sample with a thickness of 140 nm on a quartz substrate, the black appearance indicating its strong broadband absorbance.

**Photoemission Efficiency of Nanoporous Gold Photoelectrodes**

The photoresponse of npAu was evaluated using controlled-potential chronoamperometry with chopped illumination before and after coarsening through CV scans. The measurements were conducted at a potential of 0.33 V vs. RHE, where no gold oxide layer was present on the surface. Each interval of illumination generates a distinct current, as can be seen in Fig. 3(a). The difference between the current observed under illumination (at a wavelength of 291 nm) and in the dark was identified as the photocurrent ($I_{ph}$). The injection of hot electrons into the solution induced by light leads to a cathodic photocurrent. Additionally, a transient current was observed before the current saturated following the illumination. This transient current might be of capacitive nature, arising from charges trapped in surface states induced by the incident light,[44] schematic pictures of different areas in the photocurrent are shown in Fig. S6.

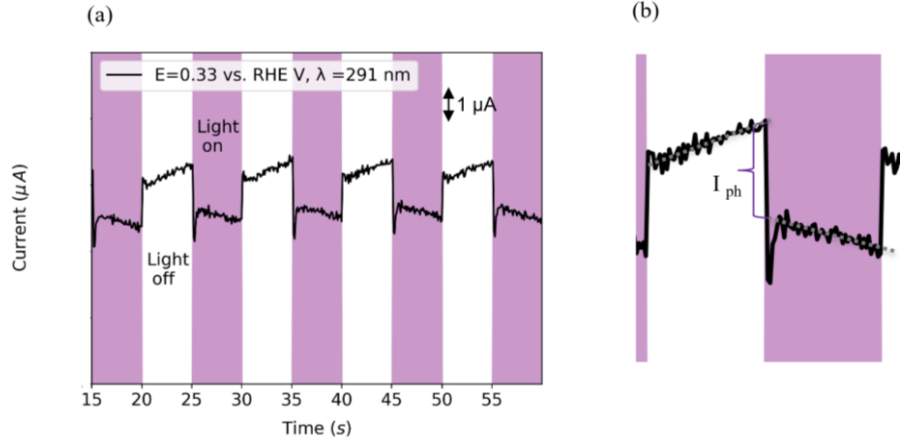

**Figure 3.** (a) Representative current transients for 140 nm-thick npAu measured at the applied potential of 0.33 V vs RHE and wavelength of 291 nm. Light illumination time was 5s. The illuminated areas are highlighted in purple. (b) Determination of the photocurrent in one illumination period.

Linear regression lines were fitted to the obtained current profiles in both the dark and illuminated areas after the transient current. The difference between the dark and light lines in the time of transition yielded the photocurrent value ($I_{ph}$), as can be seen in Fig. 3(b). These values were averaged over five pulses to obtain $I_{ph}$ and its standard deviation. The negative photocurrent ($I_{ph}$) observed in this measurement indicates electron injection into the electrolyte. The photocurrent behavior of npAu was investigated at 0.33 V vs. RHE in the pristine sample and upon coarsening via CV scans. The equilibrium potential for hydrogen evolution reaction (HER) in the dark was found to be around −0.7 V vs. RHE and independent on the ligament size,[7] as has been discussed in the coarsening section, we avoided this potential to preserve the structure.

The photocurrent $I_{ph}$ can be used to calculate the photoemission quantum efficiency. The internal quantum efficiency $\eta$ was determined to investigate the impact of ligament diameter $L$ on the photoemission efficiency:

$$\eta = \frac{N_e}{N_{ph}} = \frac{I_{ph}}{e} \frac{\hbar\omega}{AP} \tag{1}$$

$\eta$ represents the ratio of the number of emitted electrons ($N_e$) to the number of absorbed photons ($N_{ph}$). $N_e$ is related to the photocurrent $I_{ph}$ using $I_{ph} = N_e \times e$. In this equation $e$ represents the elementary charge, $A$ the absorptivity of the npAu (obtained from Fig. 2), and $P$ the incident optical power. Due to strong absorptivity of npAu the external quantum efficiency is close to internal quantum efficiency.

Figure 4 demonstrates a decrease in the total internal quantum efficiency ($\eta$) as the ligament diameter ($L$) increases from 8.4 ± 3 to 17.8 ± 5.8 nm. The obtained $\eta$ values are plotted versus $1/L$, which is fitted with a parabolic curve $b/L + c/L^2$, where we obtain $b$ equal to 0.3% nm and $c$ equal to 6.1% nm$^2$. Thus, at a ligament diameter of 10 nm, the contribution from linear term (0.03%) is smaller than that from quadratic term (0.06%).

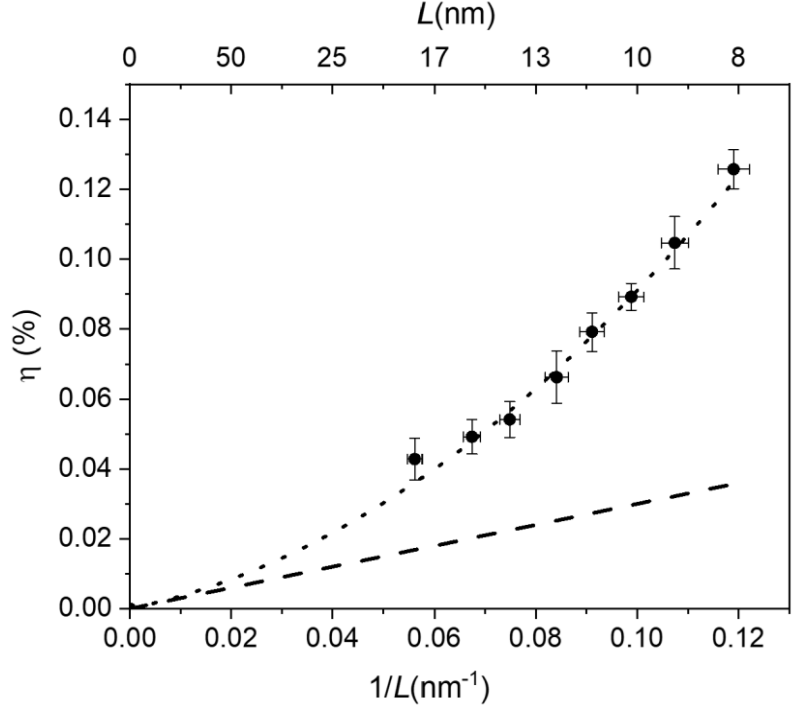

**Figure 4.** Size dependence of the internal quantum efficiency η of photoemission at a wavelength λ = 291 nm measured at 0.33 V vs. RHE. The dashed line represents a parabolic fit with linear and quadratic terms versus inverse ligament size (L$^{-1}$). Error bars were derived from the distribution width of the photocurrent Iph obtained from the repetition of the illumination pulses. The linear contribution of inverse ligament size to the total internal quantum efficiency is shown as a dashed line.

We explain the ligament size dependency and the observed quantum efficiency by splitting the quantum efficiency $\eta$ into the excitation $\eta_{ex}$ and emission $\eta_{em}$ contributions:

$$\eta = \eta_{ex}\eta_{em} \qquad (2)$$

We assume that all emitted electrons stay in water until participation in a reaction and thus contribute to the photocurrent.

First, we discuss the excitation of hot electrons. In the case of UV illumination with 4.26 eV (291 nm) photon energy, the light is absorbed by both 6sp and 5d electrons of gold. We expect that photon absorption by 5d electrons does not contribute to photocurrent as their energy is approx. 2 eV below the Fermi level[45] and they need to overcome a potential barrier larger than 3 eV above the Fermi level to be injected into water.[7,16] The participation of 6sp and 5d electrons to the photon absorption can be compared by their contribution to the imaginary part of complex permittivity $\varepsilon''$. Thus, the efficiency of 6sp electron excitation is:

$$\eta_{ex} = \frac{\varepsilon''_{6sp}}{\varepsilon''_{5d}+\varepsilon''_{6sp}} \tag{3}$$

where the 5d electron contribution $\varepsilon''_{5d} \approx 6$ at 4.26 eV.[46] The 6sp contribution can be estimated from the Drude model of free electrons, which at large frequencies can be approximated as $\varepsilon''_{6sp} = \frac{\omega_p^2 \gamma}{\omega^3}$, where $\omega$ is the frequency of light, $\omega_p$ is the plasma frequency and $\gamma$ is the collision frequency of 6sp electrons. Collision frequency consists of volume $\gamma_V$ and surface $\gamma_S$ contributions, where the volume contribution is not dependent on ligament size as long as the total number of gold atoms remains constant (which is the case here) while the surface contribution scales with $1/L$.[47,48] Frequencies can be converted into energies by multiplication with the reduced Planck constant ($\hbar$) for comparison. We use a plasma frequency of 8.5 eV,[46] volume collision frequency of 0.17 eV (0.07 eV frequency-independent volume contribution plus 0.1 eV frequency-dependent electron-electron scattering),[47] and surface collision frequency of 0.02 eV for 10 nm ligament size. We estimate the collision frequencies from the experimental results on cylindrical nanorods and nanospheres.[21,49] In this way, the estimated surface collision frequency is much smaller than assumed in our previous work.[7] With this collision frequency, the contribution of free electrons to the imaginary permittivity is $\varepsilon''_{6sp} \approx$ 0.18 at 10 nm ligament size, not changing significantly in the range of the investigated ligament sizes. Thus, the efficiency of 6sp electron excitation is 2.9% at 10 nm ligament size, of which only approx. 10% are excited by surface collisions. Taking into account that $\varepsilon''_{6sp}$ is much smaller than $\varepsilon''_{5d}$ and assuming the excitation of 5d electrons to be independent of the surface-to-volume ratio, thus, of $1/L$, the excitation probability can be split into a constant and a $1/L$ part, corresponding to the electrons excited in the volume and at the surface.

It should be noted that we assume that under illumination one photon absorbed under the condition of an electron-obstacle collision creates one hot electron.[48] However, in the UV range, there is also significant absorption by electron-electron collisions via the Umklapp process.[47] These electrons split the photon energy and thus have much lower probability to surpass the potential barrier for emission.[48]

The excited hot electrons have a finite probability of emitting into the environment $\eta_{em}$. The probability of this emission is usually expected to be rather low as hot electrons need to surpass a potential barrier.[7,15] Three effects reduce the efficiency: i) electrons excited with energy below the barrier, ii) momentum mismatch of electrons in the gold and in the conductance band of the environment, iii) reflection from the barrier. For emission into aqueous electrolyte, this results in the known scaling of efficiency with photon energy above the barrier height $n_{em} \sim (\hbar\omega - E_B - eU)^{5/2}$.[16,19] where an applied potential $U$ can shift the Fermi level position in respect to the effective conduction band of water and thus can change the effective barrier height. The emission from the npAu approximately follows this spectral relation (see Fig. S7 in Supplementary information). The estimated emission probability in this case would be in the order of 0.1% for photon energies of 1eV above the barrier height. [19,50] At the same time, increasing the ligament size should decrease the emission probability with $1/L$ dependence, as hot electrons have a lower probability of reaching the surface or interacting with it multiple times before decay. This factor is applicable to hot electrons excited by both volume and surface collisions.

The relationship between the surface area to volume ratio (*S/V*) of npAu and the reciprocal of its ligament diameter *(1/L)* is linear, given by the expression *S /V = m /L,* where *m* is a proportionality constant. For cylindrical ligaments without junctions *m* is equal to 4. The npAu has smaller surface to volume ratio for the same *L*. However, the value of *m* seems to depend on the sample preparation process and ligament diameter determination procedure, with reported values of 1.6 [51] and 2.56. [52]

Thus, we expect the internal quantum efficiency to have a contribution from the volume with $1/L$ and from the surface with $1/L^2$. From experimental results in Fig. 4, we see that the quadratic term makes a dominant 67% contribution to the efficiency at 10 nm ligament size. This is in strong contrast to the fact that only 10% of hot electrons are excited by surface collisions at this ligament size. It means that the emission probability of the surface excited hot electrons is approx. 19 times larger than that of hot electrons excited in the volume. Additionally, the measured internal quantum efficiency is strongly reduced by 5d electron absorption and the low excitation probability of 6sp electrons of 2.9%, as discussed previously. If this is taken into account, then for 10 nm ligament size with internal quantum efficiency of 0.091%, the estimated emission efficiency of all excited 6sp electrons is at 3.14%. Equation 3 can be considered with both surface and volume contributions:

$$\eta = \eta_{ex}^S n_{em}^S + \eta_{ex}^V n_{em}^V \qquad (4)$$

By knowing the quantum efficiency from the experimental results in Fig. 4 and the calculated excitation efficiency, we can estimate the emission probability of the surface and volume electrons. In this way, the probability of the surface excited electrons reaches 23%. Figure 5 depicts a diagram that summarizes the contribution of the npAu electrons to the excitation emission and the quantum efficiency according to our model. This result supports other experimental observations of the double-digit electron transfer efficiency from metal nanoparticles, where smaller barrier height and excitation with photon energies below d-band transition were employed.[53, 54] The explanation for the high hot electron emission efficiency, much higher than the theoretically expected sub percent efficiency, is a topic of ongoing debate. The correct description of the metal-semiconductor interface and contributions of possible states at the interface are discussed.[55-57]

It should also be emphasized that with the nanoporous electrode, we reach the experimental efficiencies in the order of 0.1%, several times larger than the emission observed at similar wavelengths from flat gold electrodes (Note 9 in Supplementary Information).[58] These findings further support the dominance of the surface photoelectron emission mechanism in the small ligament sizes, as predicted by theory.[59]

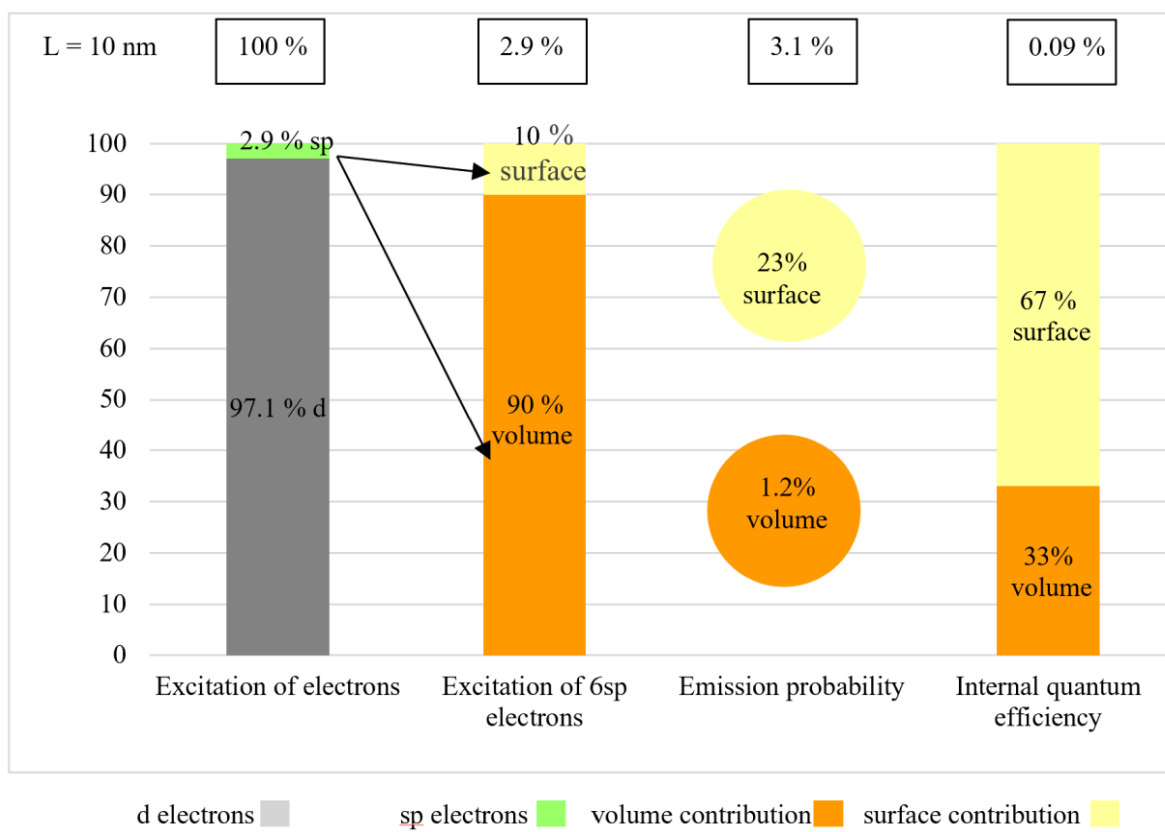

**Figure 5.** Diagram of the estimated excitation, emission, and measured internal quantum efficiency of 6sp electrons with highlighted surface and volume contributions. Numbers at the top shows the total probability of each process. The calculation was done for 10 nm ligament size.

## CONCLUSION

Our study of photoemission employs the in-situ coarsening of the ligament structure in nanoporous gold through electrochemical cycling. Using electrochemical potential cycling with a scan rate of 10 mV/s, we successfully coarsened the ligaments and decreased the surface-to-volume ratio of the npAu in small steps. We controlled the coarsening process by the upper limit of the cycling potential window and the number of cycles while performing the photoelectrochemical measurements. This method allowed us to accurately manipulate the npAu structure without changing other experimental conditions. With that approach, we have demonstrated the dominant contribution of the excited electrons from the surface collisions in photoemission.

Our results highlight the significant impact of the structure size on the internal quantum efficiency of photoemission from gold nanostructures. We observed a continuous and nonlinear decrease in the quantum efficiency for the ligament size increasing from 8 to 16 nm. Taking into account that 6sp electrons constitute only about 2.9% of all excited electrons, the obtained maximal efficiency around 0.1% is significant. Thus, double digit photoemission efficiencies can be expected from 10 nm nanostructures if excitation is performed outside of the d-band absorption range, with further improvement for smaller ligament sizes and/or using different materials.

**Conflict of Interest**

The authors declare no conflict of interest.

**Acknowledgements**

Financial support from the German Research Foundation (Deutsche Forschungsgemeinschaft, DFG) via the Collaborative Research Center SFB 986 "Tailor-Made Multi-Scale Materials Systems: M3", Project 192346071 is gratefully acknowledged.

# Supporting Information

**Size-dependent Photoemission Study by Electrochemical Coarsening of Nanoporous Gold**

*Fatemeh Ebrahimi [a,b]\*, Xinyan Wu [a], Maurice Pfeiffer [a], Hagen Renner [a], Nadiia Mameka [c], Manfred Eich [a,b], Alexander Petrov [a,b]*

[a] Hamburg University of Technology – Institute of Optical and Electronic Materials – Hamburg, Germany

[b] Helmholtz-Zentrum Hereon – Institute of Functional Materials for Sustainability – Geesthacht, Germany

[c] Helmholtz-Zentrum Hereon – Institute of Materials Mechanics – Geesthacht, Germany

\*E-mail: Fatemeh.Ebrahimi@Tuhh.de Phone number: (+49)4042878-3896

**Supplementary Note 1: Electrochemical coarsening of nanoporous gold (npAu)**

Electrochemical cycling was used to coarse the npAu electrode. Figure S1(a) shows 30 cyclic voltammetry (CV) scans on the npAu electrode in the potential window of 0.83 - 1.63 V vs. RHE in 0.5 M $H_2SO_4$. The anodic peaks in the region between 1.35 and 1.65 V correspond to the surface oxidation of npAu electrodes, while the cathodic peak at 1.17 V results from the reduction of the oxidized Au surfaces.[1]

The CVs obtained for npAu exhibit substantial differences as compared to those of a planar single crystal and a polycrystalline gold in a similar electrolyte, particularly in terms of the anodic oxidation peaks.[2] The presence of the two distinct oxidation peaks in these regions generally indicates oxide formation on the specific crystal planes exposed to the electrolyte, namely Au {100} and Au {111} surfaces at 1.4 and 1.6 V vs. RHE, respectively. The presence of the differently oriented surfaces in npAu is influenced by various factors, including a fabrication method of the precursor alloy for dealloying as well as a pretreatment of npAu before CV measurements. In addition, the existence of the oxidation of {110}-oriented surfaces around 1.45 V has been observed in some nonporous gold structures before.[3] The {110}-oriented crystal facets were also reported in npAu, where it was shown that {110} orientation can transform into {111} and {100} crystal facets through potential cycling in a sulfuric acid solution without ligament coarsening.[4] The {110} orientation and its transformation is believed to be driven by the high density of defects in the pristine npAu, which originates from the master alloy as well as the preparation method of the electrode. In reference [4] npAu in a form of a wire was converted into a powder. This method is expected to produce in a large density of defects. In contrast, we did not observe {110}-oriented surface in the CVs of our samples, consequently the surface facet changes during cycling as can be seen in Fig. S1(a). Prior to the experiment, the initial ligament size of the npAu sample was 9.1±3.8 nm (Fig. S1(b)). Since the initial ligament size of each specific sample may vary, each sample was investigated by SEM before photoelectrochemical investigation. The ligament size was acquired from scanning electron micrographs (SEM) using a Python-based AQUAMI Image analysis software.[5] The software automatically applies image thresholding and calculates the ligament dimensions of the samples. In this method, for each sample, a single SEM image is subjected to analysis, and subsequently, the average ligament diameter is computed. AQUAMI takes the input micrograph image and generates a binary image, which is then segmented into distinct phases. A comparison of the ligament size obtained by AQUAMI to the manually determined values revealed identical results.[5] Although utilizing the SEM image data containing 3D information within AQUAMI and conducting analyses akin to 2D cross-sections might not be the ideal approach, it has been commonly employed in other studies on npAu in literature.[6]

After applying 30 CV scans on the npAu, the structure coarsened so that the ligament size reached 28.9 ± 9.5 nm. In Fig. S1(b), SEM images of npAu, obtained after 2, 15, and 30 scans of potential cycling within the range of 0.83-1.63 V vs. RHE are shown.

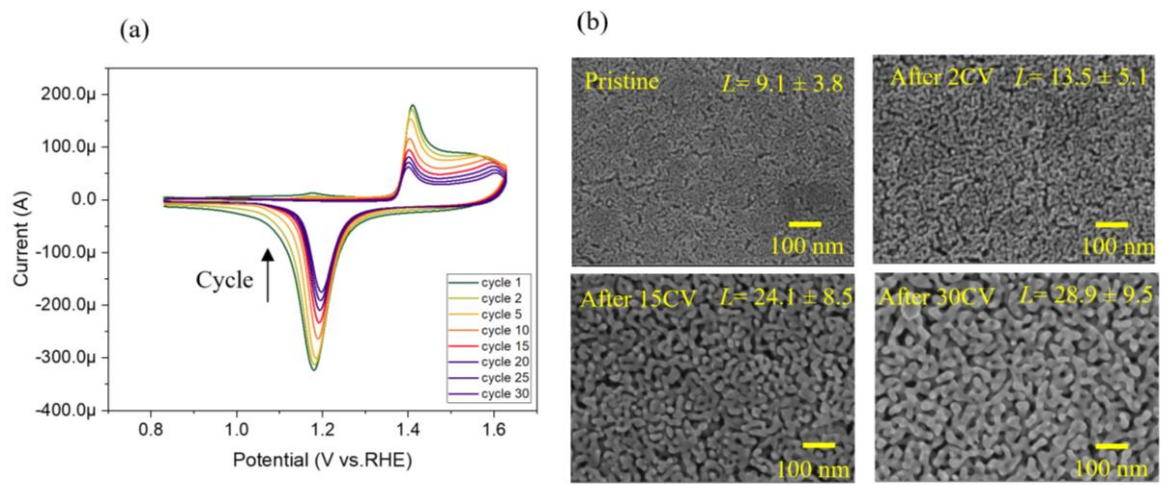

Figure S1. (a) Cyclic voltammograms of the npAu in the potential window of 0.83 - 1.63 V vs. RHE; electrolyte was 0.5 M $H_2SO_4$ and the scan rate was 10 mV/s. The area under the current reduction peak decreases with potential cycling. (b) Top-view scanning electron microscopy micrographs of the pristine sample and after the specified number of potential cycles.

**Supplementary Note 2: Surface area determination of nanoporous gold photoelectrodes using cyclic voltammetry**

Cyclic voltammetry is a widely used electrochemical technique that plays a crucial role in investigating electron transfer mechanisms. In the case of gold, determining the active surface area involves analyzing the metal oxide reduction peak, as the formation of a metal oxide is confined to a two-dimensional layer for potentials below the region of oxygen evolution. Therefore, by examining the metal oxide reduction peak, it is possible to quantify the electrochemically active surface area $A_{ECSA}$ of the npAu electrode:

$$A_{ECSA} = \frac{Q}{Q_{ox}^*} = \frac{\frac{1}{\nu}\int_{E1}^{E2}(I(E) - i_{bg})dt}{Q_{ox}^*} \tag{S1}$$

where $Q$ denotes the charge associated with the reduction of the gold oxide, $Q_{ox}^*$ is the charge that corresponds to an oxide layer that consists of one O per Au atom, $\nu$ is the scan rate of the cycling, $i_{bg}$ is the background current and $I(E)$ is the net current. It has been determined that the oxide layer containing one oxygen atom per gold atom has a charge of $Q_{ox}^*$ = 386 µC/cm².[7,8] The method can be used for flat as well as finely dispersed gold as it can be assumed that the specific adsorption on the gold surface is the same.[9]

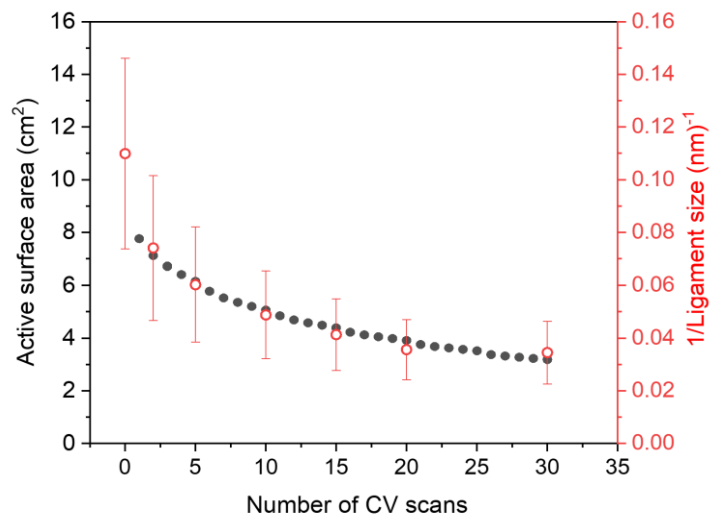

**Figure S2.** Electrochemically active area of npAu versus number of CV scans in a potential range 0.83 -1.63 V vs. RHE in 0.5 M $H_2SO_4$ along with the corresponding inverse ligament size, 1/*L*, estimated using scanning electron micrographs. The sample had a thickness of 120 nm and a macroscopically apparent surface area of 28.26 mm².

The evaluation of the electrochemically active surface area was performed based on the transferred charge during the CV measurements within the potential limits of $E_1$ = 1.4 V and $E_2$ = 0.9 V vs. RHE. $A_{ECSA}$ calculated by this method for each cycle is shown in Fig. S2 along with the inverse of ligament size, $1/L$. The decrease of the determined active area over cycling testifies to the reduction of the effective surface area of npAu electrode. A stepwise decrease in the active surface area was observed during cycling between the potentials of 0.83 and 1.63 V vs. RHE.

**Supplementary Note 3: Capacitance determination of nanoporous gold photoelectrodes using electrochemical impedance spectroscopy (EIS)**

When the oxidation-reduction reaction on the gold surface is not complete, the area under the reduction peak cannot be used to evaluate the active surface area. Therefore, for the limited potential window an electrochemical impedance spectroscopy (EIS) measurement was used to estimate the surface area of npAu based on the capacitance ratio method[10] (see Supplementary Note 4 below). Both methods, the capacitance ratio and oxygen adsorption from solution, explained in Note 3, have their limitations and can be considered to be only used to estimate the surface area. Both methods are not expected to result in the same value of the real surface area of the electrodes.[11]

In order to calculate the capacitance of the npAu electrode, the EIS measurement was conducted on the npAu electrode at fixed value of a DC potential (0.83 V vs. RHE) in the capacitive region to avoid any faradaic reactions and ensure only double-layer charging. The impedance spectrum was recorded from 10 Hz to 100 Hz with the logarithmic distribution per decade using an AC potential of 14 mV peak-to-peak. For a given electrode, in the absence of faradaic processes, the EIS measurement results are interpreted via an electronic circuit containing one resistor (R) in series with a capacitor (C), where both elements represent the ohmic solution resistance and the capacitance of the double layer, respectively. Thus, the total impedance of the metal-solution interface can be expressed as $Z = R - i\,(1/\omega C)$.[10] The capacitance is calculated based on the experimentally obtained imaginary part of the impedance ($Z''$) and the angular frequency of the applied AC signal according to $Z'' = 1/\omega C$. The findings of the EIS measurement is shown in Fig. S3(a), where the data points are fitted with an equation $Z'' = 1/\omega C$ so that the slope will give the capacitance. The EIS measurement was done before and after cycling, and the capacitance was calculated, as seen in Fig. S3(b).

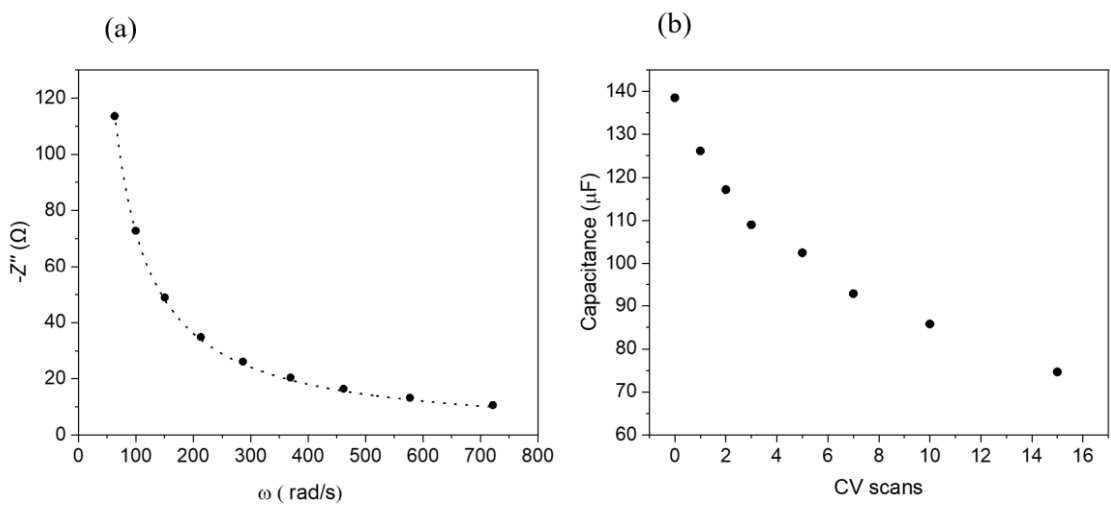

**Figure S3.** (a) Dependence of the imaginary part −Z′′ on the angular frequency ω fitted with an equation Z′′ = 1/ωC for determination of the double layer capacitance of a npAu pristine sample in 0.5 M $H_2SO_4$. (b) Capacitance of the npAu-electrolyte calculated by EIS measured at 0.83 V vs. RHE on the pristine sample and after a different number of CV scans in the potential window of 0.53 -1.43 V vs. RHE. The sample had a thickness of 120 nm and a macroscopically apparent surface area of 28.26 mm².

**Supplementary Note 4: Determination of a ligament size of nanoporous gold photoelectrodes by electrochemical impedance spectroscopy**

The ligament size can be acquired from the electrochemical surface area of the nanoporous electrode.

After the EIS measurement and the calculation of the capacitance, the electrochemical surface area npAu can be evaluated using the capacitance ratio method. This method involves estimation of the electrochemical surface area by measuring the apparent total capacitance ($C$) and comparing it to the specific double-layer capacitance (capacity per area) of the planar gold surface ($C^*$):[11]

$$A_{\text{electrochemical}} = \frac{C}{C^*} \qquad (S2)$$

Various approaches have been employed to estimate the exact relation between the electrochemical surface area and the ligament size of nanoporous structures. Often volume-specific surface area ($\alpha_v$) is used as a quantitative measure of the ligament size. One commonly used relationship is $L = 4/\alpha_v$ by assuming idealized cylindrical ligaments with a diameter of $L$.[12] In order to avoid errors originating from the relation between these two parameters, we consider a linear relation between the capacitance of npAu and the inverse ligament size $C =$ $\beta \frac{1}{L}$. To calculate the coefficient $\beta$ of this relation, we use the average ligament size of the pristine and the final state (after 15 CV scans) of the sample acquired by SEM and the capacitance resulting from the EIS. The calculated in this way coefficient $\beta$, we used that to estimate the ligament size after CV scans, assuming the constant volume of the sample during coarsening. The estimated ligament size by this method is shown in Fig. S4(a) versus the number of the CV scans applied.

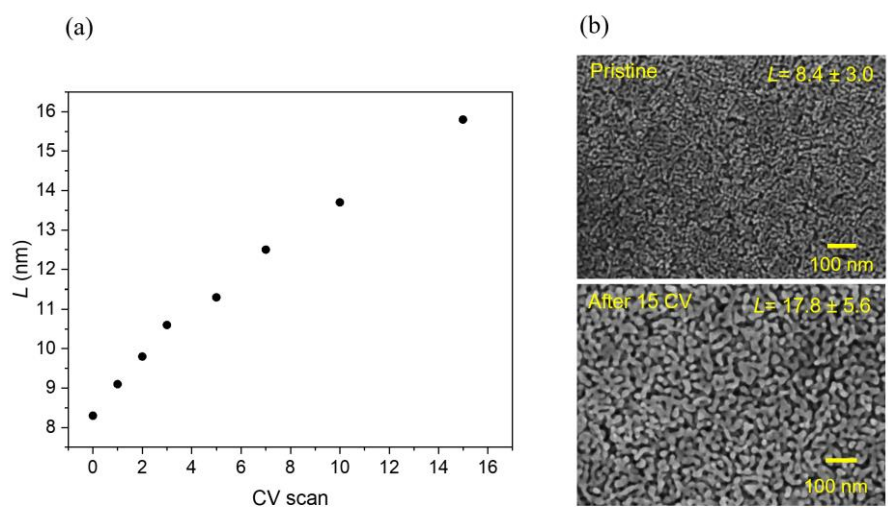

**Figure S4.** (a) Estimated ligament size using the capacitance of the npAu and SEM analysis before and after potential cycling. (b) SEM picture of pristine sample and after 15 CV scans in the potential window of 0.53 -1.43 V vs. RHE.

**Supplementary Note 5: Emission of electrons into the electrolyte**

The light-induced injection of electrons into solution leads to a cathodic photocurrent which we correlate with the photoemission quantum efficiency. Hot electrons emitted from plasmonic metal nanostructures produce solvated electrons. These solvated electrons can either recombine by going back into the electrode or be captured by other ions or molecules in the solution.[13] In our case, $H_3O^+$ ions in the $H_2SO_4$ electrolyte play a vital role as charge carriers.[14] They facilitate the capture of the emitted electrons, which is critical for the overall process. These captured electrons may potentially contribute to other reactions, with hydrogen production being a potential pathway for the hot electrons ejected from gold in the electrolyte.[15]

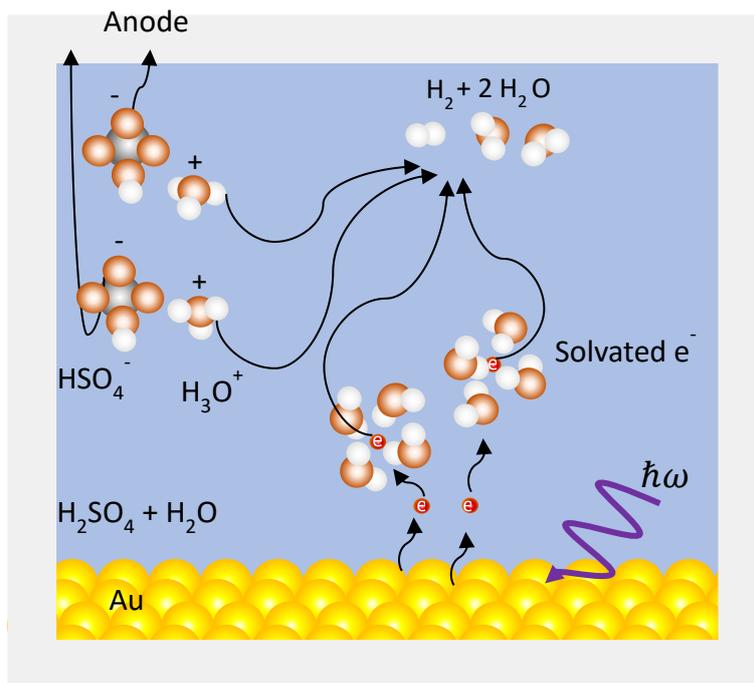

**Figure S5.** Scheme explaining the sequential steps involving solvated electrons and their subsequent capture process by $H_3O^+$.

**Supplementary Note 6: Different regions in transient photocurrent response**

After preparation, the npAu photoelectrode was mounted in the photoelectrochemical cell. Typically, the photocurrent curve exhibits two distinct regions: a rapid response current (RRC) and a slow response current (SRC) region (usually reported on a time scale of 10 s). Comparable photocurrent responses to RRC are observed in metal- semiconductor photoelectrode and gold nanostructures due to excited carriers.[16] Therefore, we believe that the presence of the excited carriers is responsible for the observed RRC in the npAu electrode. On the other hand, the SRC is associated with a photothermal processes.[17] An estimation of temperature changes in npAu is given in Note 7.

Additionally, a transient current (TC) was observed at the beginning of each illumination, occurring within a short time scale of less than 0.05 seconds. The TC is attributed to the injection of photo-generated hot carriers into surface states, which is typically observed in semiconductor systems[18] and also reported in gold nanostructures.[16]

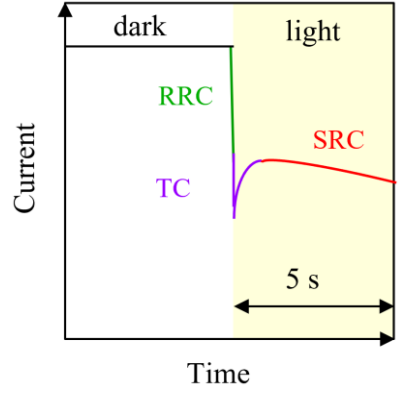

**Figure S6.** Scheme explaining the rapid current (RRC), slow response current (SRC), and the transient current (TC) regions upon illumination.

**Supplementary Note 7: Temperature dynamics in photoheated npAu sheets**

The thin npAu film heated by absorbed radiation will lose thermal energy to the surrounding water by heat diffusion. The process can be modelled as a thin film source in water. The one-dimensional heat-diffusion equation for the evolution of the temperature profile $T(t,x)$ under an illumination switched on at $t = 0$ and being $R(t) = R_0$, i.e., constant afterwards can be formulated as:

$$\frac{dT(t,x)}{dt} = D\frac{\partial^2 T(t,x)}{\partial^2 x} + R_0 F(x) \tag{S3}$$

where $t$ is the time, $x$ is the spatial direction orthogonal to the film, $D$ is the diffusion constant, $F(x) = 1$ in $-w/2 < x < w/2$ and $F(x) = 0$ elsewhere, and $w$ is the film thickness. If heat is not able to diffuse away ($D = 0$) the temperature in the npAu film would evolve as

$$T(t, -w/2 < x < w/2) = R_0 t \tag{S4}$$

where

$$R_0 = \frac{AJ}{\rho_1 c_1 w}. \tag{S5}$$

Here, $A$ is the absorptivity, $J$ the incident light intensity, and $\rho_1$ and $c_1$ are effective density and specific thermal capacity of npAu, respectively. So in the middle of the gold sheet we have initially for small times

$$T(t, x = 0) = \frac{AJ}{\rho_1 c_1 w} \cdot t, \qquad 0 \leq t \ll t_\times$$

(S9)

and asymptotically for large times

$$T(t \to \infty, x = 0) \cong \frac{AJ}{\rho c}\sqrt{\frac{t}{\pi D}}, \qquad t_\times \ll t < \infty. \tag{S6}$$

The experimental parameters of water, $c$ = 4.18 J/(g K), $\rho$ = 1 g/cm³, $D$ = 0.00143 cm²/s, $A$ = 90%, $J$ = 0.184 W/cm², within an illumination interval of $t$ = 5 s are considered. At time $t_x = w^2/\pi D$ the two approximations intersect, so far beyond $t_x$ the diffusion turns the initially linear temperature growth into a progressively slower one. Since $t_x$ is much smaller than a second the overall behavior in our experiments is dominated by eq. (S6). Thus the latter specifies the maximum temperature increase to be expected. The expected increase in temperature is around 0.04 K, which may seem negligible in most scenarios, but it carries significant implications in the context of nanoscale photoeffects. During a 5-second illumination, we observed slow response changes in current of approximately 0.02 µA. Previous studies have indicated that changes in the temperature of gold nanostructures may lead to an increase of 10 µA in the cathodic current with a temperature rise of about 10°C.[16] However, due to the influence of the temperature on the ligament sizes, direct temperature-dependent measurements were not taken in our experimental setup.

**Supplementary Note 8: Energy threshold required for the electron injection into water**

The photoemission spectra of the representative npAu electrodes revealed the energy threshold required for the electron injection into water, as shown in Fig. S7. The minimum photon energy needed to generate a detectable current signal (red edge energy) was found to be approximately 3.2 eV. It can be observed that increasing the bias potential ($E$) enhances the overall internal quantum efficiency ($\eta$) as the Fermi level of gold (Au) is elevated towards the conductance band of water. The probability of an electron leaving the electrode is proportional to the excess energy above the photoemission threshold according to previous studies:[16,19]

$$\eta_S \sim \frac{1}{L^2}(\hbar\omega - E_B - eU)^{5/2} \tag{S7}$$

Here, $E_B$ presents the barrier energy for electron injection into the water, $e$ denotes the elementary charge, $E$ is the applied potential, and $\hbar\omega$ corresponds to the incident photon energy. The dependence of $\eta$ on $\hbar\omega$ approximately fits the anticipated power law of 5/2.

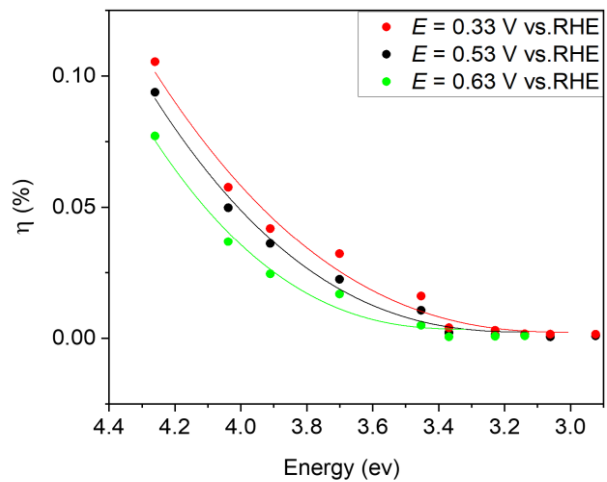

**Figure S7.** Dependence of the internal quantum efficiency of npAu on photon energy at different bias potentials with fits according to the anticipated power law of 5/2.

**Supplementary Note 9: Photoemission of planar gold**

Figure S8 shows the photoresponse of a sputtered planar gold, which was evaluated using chronoamperometry with chirped illumination. The measurements were conducted at a potential of 0.33 V vs. RHE, where no OH adsorbates present on the surface. The measured photoemmision of the planar gold is 0.021 ± 0.002 µA. Considering the absorptivity of the planar gold, which is shown in Fig. S8(b), the internal quantum efficiency of the electron emission from planar gold is 0.031± 003%, which is three times smaller than that of npAu with 10 nm ligament size.

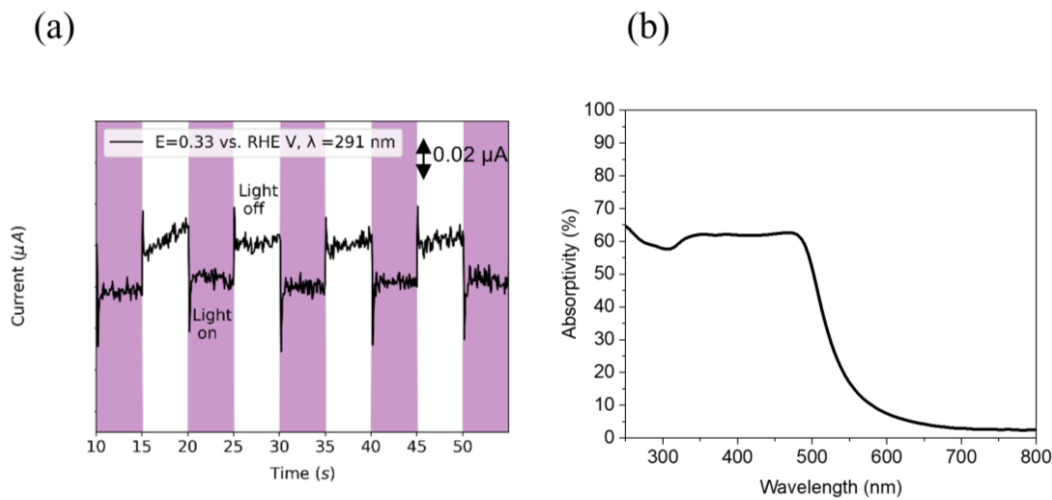

**Figure S8.** (a) Representative current transients for a planar gold measured at the applied potential of 0.33 V vs RHE and wavelength of 291 nm. Illumination time periods were 5s long, highlighted in purple (b) Absorptivity spectra of the planar Au film.